\documentclass[%
 aip,
 amsmath,amssymb,
 reprint,%
floatfix
]{revtex4-1}

\usepackage{graphicx}
\usepackage{dcolumn}
\usepackage{bm}

\usepackage[utf8]{inputenc}
\usepackage[T1]{fontenc}
\usepackage{mathptmx}
\usepackage{etoolbox}

\makeatletter
\def\@email#1#2{%
 \endgroup
 \patchcmd{\titleblock@produce}
  {\frontmatter@RRAPformat}
  {\frontmatter@RRAPformat{\produce@RRAP{*#1\href{mailto:#2}{#2}}}\frontmatter@RRAPformat}
  {}{}
}%
\makeatother

\begin{document}

\preprint{AIP/123-QED}

\title{A Novel Architecture for room temperature \\microwave optomechanical experiments}
\author{Sumit Kumar}
 \affiliation{Department of Physics, Royal Holloway University of London,\\ Egham, Surrey, TW20 0EX, U.K.}
 \author{Sebastian Spence}
 \affiliation{Department of Physics, Royal Holloway University of London,\\ Egham, Surrey, TW20 0EX, U.K.}
  \author{Simon Perrett}
 \affiliation{Department of Physics, Royal Holloway University of London,\\ Egham, Surrey, TW20 0EX, U.K.}
  \author{Zaynab Tahir}
 \affiliation{Department of Physics, Royal Holloway University of London,\\ Egham, Surrey, TW20 0EX, U.K.}
  \author{Angadjit Singh}
 \affiliation{Clarendon Laboratory, Department of Physics, University of Oxford,\\ Oxford OX1 3PU, U.K.}
  \author{Chichi Qi}
 \affiliation{Department of Physics, Royal Holloway University of London,\\ Egham, Surrey, TW20 0EX, U.K.}
  \author{Sara Perez Vizan}
 \affiliation{Department of Physics, Royal Holloway University of London,\\ Egham, Surrey, TW20 0EX, U.K.}
  \author{Xavier Rojas}
 \affiliation{Department of Physics, Royal Holloway University of London,\\ Egham, Surrey, TW20 0EX, U.K.}
 \email{xavier.rojas@rhul.ac.uk}
\date{\today}

\begin{abstract}
We have developed a novel architecture for room temperature microwave cavity optomechanics, which is based on the coupling of a 3D microwave reentrant cavity to a compliant membrane. Devices parameters have enabled resolving the thermomechanical motion of the membrane, and observing optomechanically induced transparency/absorption in the linear regime, for the first time in a microwave optomechanical system operated at room temperature. We have extracted the single photon coupling rate ($g_0$) using four independent measurement techniques, and hence obtain a full characterization of the proposed cavity optomechanical system.
\end{abstract}

\maketitle
\section{Introduction}\label{sec:level1}
The advances in low-temperature microwave optomechanics has successfully demonstrated many groundbreaking experiments such as ground-state cooling~\cite{cattiaux2021macroscopic,teufel2011sideband}, backaction evading measurements~\cite{suh2014mechanically,ockeloen2016quantum}, entangling massive mechanical resonators~\cite{ockeloen2018stabilized,kotler2021direct}, and ultrasensitive detection of small forces~\cite{weber2016force,hanay2012single}. Microwave optomechanical technologies can also foster the architecture for signal processing~\cite{kumar2022microwave,massel2011microwave} and quantum limited measurements~\cite{clark2018cryogenic,barzanjeh2017mechanical}.

The field of cavity optomechanics has also facilitated many precision sensing applications at room temperature (RT). Schliesser \textit{et al.} \cite{schliesser2008high} achieved the displacement sensitivity of $10^{-19} \text{ m}/\sqrt{\text{Hz}}$ by  coupling light and mechanical mode of a silica toroidal cavity. Sansa \textit{et al.} \cite{sansa2020optomechanical} demonstrated an optomechanical mass spectrometry able to detect mass up to 100 kDa. Gavartin \textit{et al.} \cite{gavartin2012hybrid} reached a force sensitivity of 74 aN/$\sqrt{\text{Hz}}$ by coupling Si$_3$N$_4$ nanomechanical beam to a disk-shaped optical resonator. Other applications include atomic force microscopy \cite{srinivasan2011optomechanical,allain2020optomechanical}, accelerometry \cite{krause2012high}, magnetic resonance force microscopy \cite{mamin2003detection}. Fabricating resonators with low mechanical dissipation can also facilitate quantum optomechanical experiments at RT \cite{norte2016mechanical}. However, all the above mentioned optomechanical experiments at RT exist in the optical domain only. Developing microwave optomechanics at RT can further advance the field of ultra-precision sensing and is suitable for large-scale integration \cite{faust2012microwave}.

To realize RT microwave optomechanics, one needs to optimize cooperativity \cite{serra2021silicon}. The cooperativity $C=4g_0^2n_\text{d}/(\kappa \Gamma_\text{m})$ characterizes the efficiency of the exchange of photons and phonons in an optomechanical system \cite{aspelmeyer2014cavity}, where $g_0$ is the single-photon coupling rate between the microwave cavity and the mechanical resonators' mode of interest, $n_\text{d}$ is the number of the photons inside the cavity, $\kappa$ and $\Gamma_\text{m}$ are cavity and mechanical resonator decay rates respectively. RT microwave optomechanical experiments have been previously accomplished using microstrip resonators made of normal conducting Cu coupled to a mechanical element \cite{faust2012microwave}. These resonators are marred by dielectric and conductor losses, limiting their quality factor to about 100 which adds a constraint to the displacement and the force sensitivity \cite{le2021room}.
 
One can benefit by coupling mechanical resonators to 3D microwave cavities with low loss. In a recent work by Liu \textit{et al.} \cite{liu2022quantum}, backaction evading measurements of a silicon nitride membrane resonator were performed via coupling it to a 3D Cu cavity. They observed a cavity quality factor of 15,000 at 300 mK. Yuan \textit{et al.} \cite{yuan2015large} were able to achieve optomechanical cooperativity of 146,000 by coupling silicon nitride membrane to a three-dimensional superconducting microwave cavity. However, both experiments required low temperatures and an antenna chip to be flip-chip bonded to the SiN membrane resonator for microwave coupling, increasing fabrication steps. Tuan \textit{et al.} \cite{le2021room} also employed a 3D microwave cavity made of Cu to dielectrically couple to a Si$_3$N$_4$ string. They characterized the optomechanical coupling by driving the string using the two tones in a non-linear regime at RT. Pearson \textit{et al.} \cite{pearson2020radio} also characterized the mechanical properties of driven Si$_3$N$_4$ membrane using a radio-frequency 2D cavity at RT.
 
Here, we present a novel architecture for room temperature microwave optomechanics. Our cavity optomechanical system is composed of a Si$_3$N$_4$ membrane coupled to a 3D microwave re-entrant cavity. Our relatively high cooperativity for a microwave optomechanical system enable detecting thermal motion and optomechanically induced transparency/absoroption (OMIT/OMIA) with great resolution. Using four independent characterization techniques, we extract the single photon coupling rate $g_0$, a key metric for cavity optomechanical systems. We pushed this system into the nonlinear regime to extract the duffing parameter, and hence defining the dynamic range for sensing applications. 

\section{Optomechanical system}
\begin{figure}[b]
	\includegraphics[width=\columnwidth]{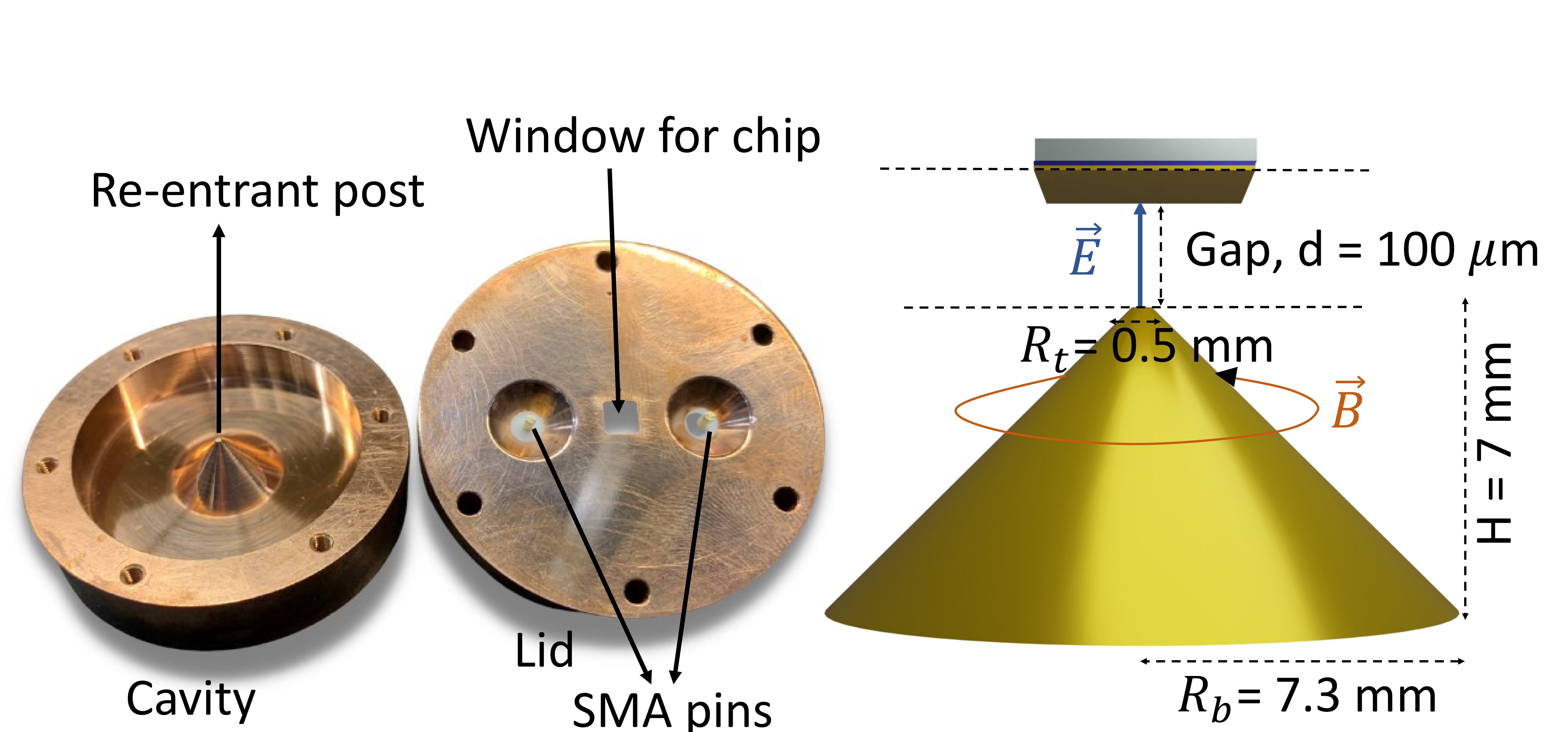}
\caption{\label{schematic} (left) The cavity with chip window and SMA pin couplers. (right) Schematic of the re-entrant pillar and Si$_3$N$_4$ membrane chip. The cylindrical cavity with an inner radius of $R_\text{cav}=$ 18 mm with a conical post of height $H=$ 7 mm, bottom radius $R_\text{b}=$ 7.3 mm and top radius $R_\text{t}=$ 0.5 mm constitutes our 3D cavity. The electric field of the fundamental mode of the cavity is confined between the chip and top of the post whereas the magnetic field is around the post }
\end{figure}

The cavity optomechanical systems consists of a Si3N4 membrane coupled to a 3D microwave cavity. The microwave re-entrant cavity consists of a post enclosed in a cylindriccal cavity leaving a tiny gap between the top of the post and a cavity wall, as shown Fig. \ref{schematic}. The cavity is made of Cu which is electropolished and gold plated. The electric field of the fundamental electromagnetic mode within is highly confined within the gap between the top of the post and the wall of the lid, which for this experiment is a window to a gold plated Si$_3$N$_4$ membrane, as shown in Fig. \ref{schematic}.

The resonance frequency of the cavity is given by $\omega_\text{c}=1/\sqrt{LC}$. The capacitance $C = C_0+C_1$ where $C_0$ is the capacitance between the top of the post and the wall of the lid determined by the gap and $C_1$ is the capacitance of the rest of the cavity. The inductance $L$ is defined by the geometry of the post inside the cavity. Any change in the gap will be translated to a change in the resonance frequency of the cavity. Thus, if the end wall is made of any mechanical resonator, the system behaves as a transducer for detecting mechanical motion. We incorporated a 5mm $\times$ 5mm chip with a mechanical resonator within a window in the lid. The chip is placed across the window from the reverse side of the lid and secured in place with a PTFE grub screw. This eliminates the need for adhesives which can reduce the cavity quality factor. 

The mechanical resonator is a 2 mm $\times$ 2 mm pre-stressed Si$_3$N$_4$ membrane \cite{Norcada}. The Si$_3$N$_4$ membrane is 50 nm thick, plated with 55 nm of Au. The motion of the mechanical resonator will thus induce modulation of the resonance frequency of the cavity owing to a change in the gap. The optomechanical coupling strength $G=\partial \omega_\text{c}/\partial x$ is inversely proportional to the gap $x$. We targeted the gap to be $x \sim$ 100 $\mu$m when machining the cavity. We then estimated the gap to be close to $x\sim$ 100 $\mu$m (see Appendix \ref{gap_estimation}) by comparison with the analytical expression for the resonance frequency of the re-entrant cavity and the FEM simulation using COMSOL. We attached a piezo-electric transducer (PZT) to the cavity lid. The cavity is placed inside a vacuum chamber with pressure below 5 $\times$ $10^{-5}$ mbar. The pressure gauge located on the vacuum chamber did not work below this value. However, our pumping system should easily enable us reaching the 10$^{-6}$ mbar range. All the measurements were performed at room temperature.          

\section{Characterization}
The two port microwave cavity is characterized via measurement of scattering parameters with a vector network analyzer (VNA). The external microwave feedlines are capacitively coupled to the cavity using SMA pins. The length these pins protrude into the cavity determines the coupling strength $\kappa^1_\text{ext}$ and $\kappa^2_\text{ext}$ of port 1 and port 2, respectively. The dielectric and conductor losses are characterized by $\kappa_\text{int}$ such that the total loss in the cavity is $\kappa=\kappa^1_\text{ext}+\kappa^2_\text{ext}+\kappa_\text{int}$. From our measurements, the resonance frequency of the cavity is $\omega_\text{c}/(2\pi)$= 4.537 GHz, the linewidth is $\kappa/(2\pi)$= 5.504 MHz, and external linewidths are $\kappa^1_\text{ext}/(2\pi)$=$\kappa^2_\text{ext}/(2\pi)\approx$ 245 kHz. The amplitude $|S_{21}|$ is plotted in the Fig. \ref{figure_2}. The quality factor of the cavity is 800 and the coupling efficiency is $\kappa^1_\text{ext}/\kappa$= 0.044, showing the cavity is under-coupled. The quality factor can be further increased by surface treatment of the Cu. 

The characteristics of the fundamental mode of the mechanical resonator and the single photon optomechanical strength $g_0$ were extracted using four different methods. Firstly, we measured the thermomechanical motion by driving the cavity at its resonance frequency $\omega_\text{c}$ and measuring the noise spectrum at $\omega_\text{c}+\Omega_\text{m}$ where $\Omega_\text{m}$ is the resonance frequency of the fundamental mode of the mechanical resonator using the spectrum analyzer. Secondly, we again measured the noise spectrum at $\omega_\text{c}+\Omega_\text{m}$ by driving the cavity at $\omega_\text{c}$ and the mechanical resonator using white noise imparted by the PZT. Thirdly, we characterized $g_0$ using two-tone measurements thus demonstrating OMIT/OMIA in the response of the cavity and at last, we employed a homodyne detection scheme to extract $g_0$. Finally, we study the mechanical resonator in the nonlinear regime to extract the duffing parameter, and hence quantify the dynamic range for potential sensing applications.

\subsection{\label{subsec:level1} Thermal noise measurement}
\begin{figure}[b]
	\includegraphics[width=\columnwidth]{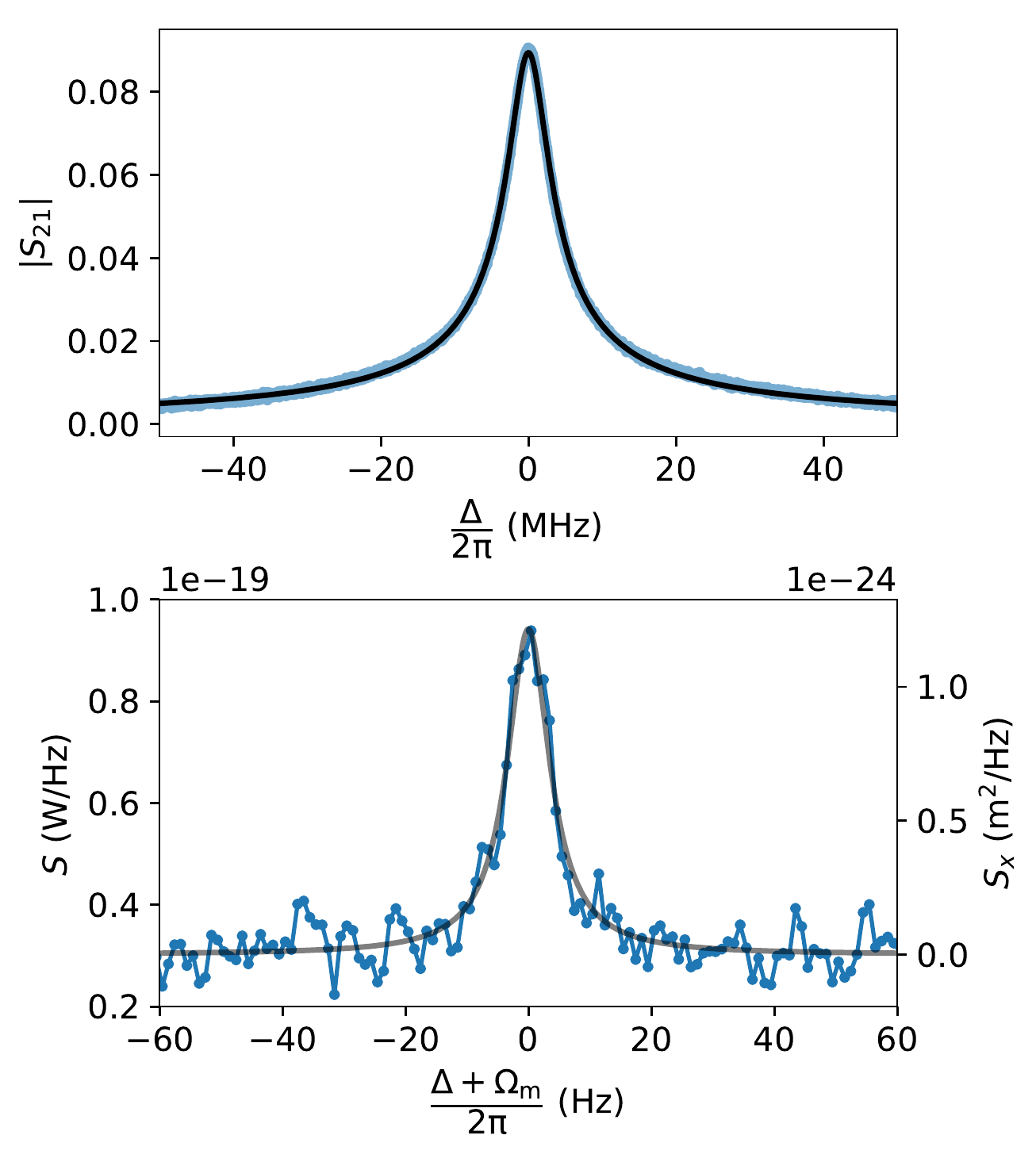}
	\caption{\label{figure_2} (top) The transmission S$_{21}$ is plotted for the microwave cavity. (bottom) Noise power spectral density of mechanical sideband at $\Delta+\Omega_\text{m}$ due to brownian motion of Si$_3$N$_4$ membrane. For both the plots, experimental data is shown in blue and the fits are in grey. Noise floor has been subtracted on the right scale.
	}
\end{figure}

The motion of the mechanical resonator with resonance frequency $\Omega_\text{m}$ will modulate the cavity drive frequency $\omega_\text{d}$ generating sidebands at $\omega_\text{d} \pm \Omega_\text{m} $. The thermomechanical motion of the Si$_3$N$_4$ membrane was measured by applying a microwave signal to port 1 of the cavity at $\omega_\text{d}$ and measuring the noise spectrum of the transmitted signal at $\omega_\text{d} + \Omega_\text{m} $ using a spectrum analyzer. The loss and gain contributions of the input and the output lines were removed via careful calibrations. The measurement circuit diagram is described in appendix \ref{circuit diagram}. We focus on the fundamental flexural mode with resonance frequency $\Omega_{11}$ of the Si$_3$N$_4$ membrane. Theoretically, the resonance frequency $\Omega_{11}$ can be determined using the square membrane formula \cite{hauer2013general},
 \begin{equation}
 	\dfrac{\Omega_{11}}{2\pi} = \dfrac{1}{2}\sqrt{\dfrac{\sigma}{\rho}\dfrac{2}{a^2}}\text{ ,}
 \end{equation}
 where $\sigma=$ 1 GPa is the stress in the Si$_3$N$_4$ membrane (information provided by the supplier Norcada Inc. \cite{Norcada}), $\rho=2820\pm160$ kg/m$^3$ is the mass density of the Si$_3$N$_4$, and $a=$ 2 mm is its lateral dimension. 
However, the resonance frequency changes due to the loaded mass of 55 nm Au on the membrane. The modified resonance frequency $\Omega_\text{m}=2\pi f_{11}$ is given by \cite{cleveland1993nondestructive},
\begin{equation}
	f_{11}=\dfrac{1}{2\pi}\sqrt{\dfrac{m_\text{eff,11}\Omega_\text{11}^2}{m_\text{eff,11}+m_\text{load}}}\text{ ,}
\end{equation}
where $m_\text{eff,11}$ is the effective mass of the fundamental flexural mode of Si$_3$N$_4$ membrane and $m_\text{load}$ is the effective mass of the loaded Au on Si$_3$N$_4$ membrane. Theoretically, $f_{11}=$ 72 kHz. The cavity mode is pumped by applying a microwave signal $\omega_\text{d}$. We also define a detuning $\Delta = \omega_\text{d}-\omega_\text{c}$. For in-cavity pumping, the number of drive photons $n_\text{d}$ inside the cavity is given by \cite{kumar2021low},
 \begin{equation}
 	n_\text{d}=\dfrac{2\kappa_\text{ext}P_\text{in}}{\hbar \omega_\text{c} (\kappa^2+4\Delta^2)}\text{ ,}
 \end{equation}
where $\kappa_\text{ext}=\kappa_\text{ext1}+\kappa_\text{ext2}$ and $P_\text{in}$ is the power incident on port 1. The number of photons $n_\text{d} \sim$ $4.1 \times 10^{11}$ are present in the cavity for 5$\pm$0.5 dBm (depending on 0.5 dBm of error made on the calibration of the power) of power applied by
the microwave generator which is attenuated by 11.8 dB before reaching the input port of the cavity. The power spectral density, $S[\Omega]$ (Watts/Hz) at $\Omega_\text{d}  + \Omega_\text{m} $ coming out of port 2 of the cavity in the frame rotating with $\omega_\text{c} $ is given by \cite{kumar2021low,zhou2019chip},
\begin{equation}{\label{equation1}}
	S[\Omega](\text{Watts/Hz})= \dfrac{\kappa_\text{ext}G^2 n_\text{d}}{\kappa^2}\hbar\omega_\text{c} S_{x} [\Omega] + \text{nf}\text{ ,}
\end{equation}
where nf is the noise floor and $S_\text{x}[\Omega]$ is given by,
\begin{equation}{\label{equation4}}
	S_{x} [\Omega] (\text{m}^2/\text{Hz}) = \dfrac{4 k_\text{b} T}{m_\text{eff}\Omega_\text{m}^2}\dfrac{\Gamma_\text{m}}{\Gamma_\text{m}^2+4(\Omega-\Omega_\text{m})^2}\text{ ,}
\end{equation}  
  is the single-sided noise position spectral density \cite{pernpeintner2016nanomechanical}. $k_\text{b}$ is the Boltzmann's constant, $T=$ 293 K, $\Gamma_\text{m}$ is the mechanical linewidth, $x$ is the global RMS displacement of the Si$_3$N$_4$ membrane normalized by mode shape function at temperature $T$ and $\Omega$ is close to $\Omega_\text{m}$. The power spectral density at $\Delta+\Omega_\text{m}$ is plotted in the Fig. \ref{figure_2}. From a Lorentzian fit of the experimental data, we extracted $\Omega_\text{m}/(2\pi)=$ 71.4 kHz and $\Gamma_\text{m}/(2\pi)=$ 8.4 Hz, giving rise to the mechanical quality factor $Q$ of 8500.
  
  Next, we focused on extracting single-photon coupling rate $g_0= Gx_\text{zpf}$, where $x_\text{zpf}=\sqrt{\hbar/(2m_\text{eff}\Omega_\text{m})}$ is the zero-point fluctuation of the mechanical mode and $m_\text{eff}=m_\text{eff,11}+m_\text{load}$ = 1.20 ng giving $x_\text{zpf}=$ 0.31 fm. The noise floor nf is subtracted from the power spectral density $S$ such that $S-\text{nf}$ is written in terms of position spectral density $S_{x}$. We then fitted the resultant plot using Eqs. \ref{equation1} and \ref{equation4}. If we take the theoretical value of $m_\text{eff}$ and $x_\text{zpf}$, only $g_0$ remains unknown. From the fit, we extracted $g_0/2\pi= G x_\text{zpf}/2\pi=$ 185$\pm 5$ $\mu$Hz.
 
  The noise floor in our detection signal corresponds to a displacement sensitivity of 0.75 pm/$\sqrt{\text{Hz}}$. This represents approximately an order of magnitude improvement from previously reported displacement sensitivity \cite{faust2012microwave}.

\subsection{\label{subsec:level2}White noise drive}
\begin{figure}[ht!]
	\includegraphics[width=\columnwidth]{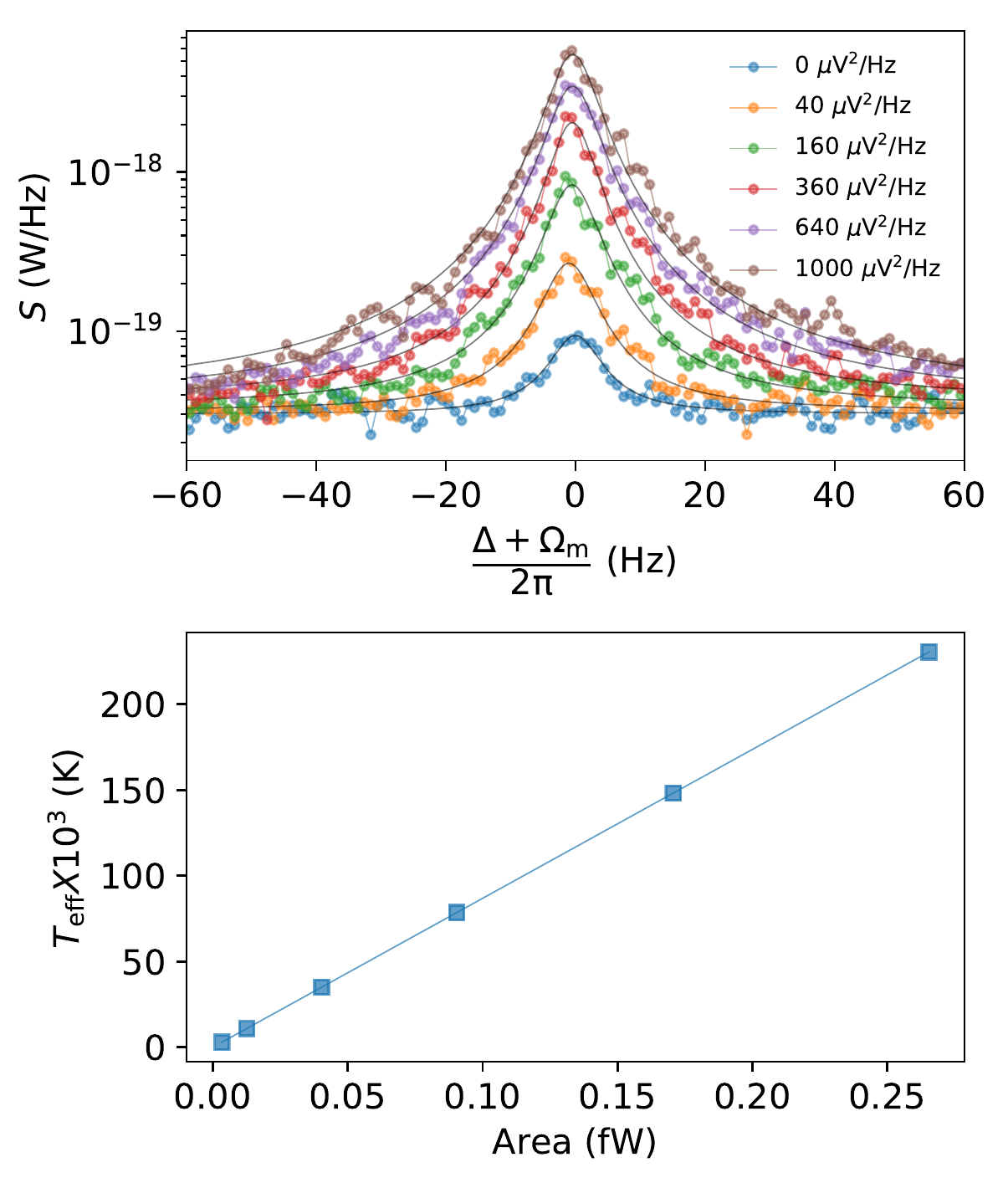}
	\caption{\label{figure_3} (top) The power spectral density of mechanical sideband at $\omega_\text{c}+\Omega_\text{m}$ is plotted for different values of white noise spectral density imparted by PZT drive. The fits are shown in grey. (bottom) The effective temperature is plotted versus the area under power spectral density. The slope $\zeta$ of the linear data is used in Eq. \ref{equation2} to extract $g_0$. 
	}
\end{figure}

Extracting $g_0$ is usually a difficult task unless one can directly measure the thermal motional sidebands as described in the previous section. In some cases, there is not enough sensitivity to detect the thermal motion of the mechanical resonator. One can therefore drive the mechanical system with an external force as the one generated from, for instance, a piezoelectric transducer. However, it is difficult to know precisely the conversion factor between the voltage applied across the piezoelectric transducer and the force generated on the mechanical resonator, hence preventing the derivation of $g_0$. Instead, the technique we present now uses a source of white noise to drive the piezoelectric transducer, generating a noise force on the mechanical resonator. This increases the effective mode temperature and the amplitude of the mechanical sideband measured at $\Delta+\Omega_\text{m}$. Fig. \ref{figure_3} shows the mechanical sideband at $\Delta+\Omega_\text{m}$ when the cavity is pumped at $\omega_\text{c}$ for different RMS voltages of white noise applied to the transducer. One can see roughly two orders of magnitude of increase in the amplitude of the mechanical sideband when driven by 1000 $\mu$V$^2/$Hz of white noise generated by the noise source in the bandwidth of 10 MHz compared to the undriven case. From Eq. \ref{equation1}, the area under the power spectral density is directly proportional to the temperature bath. We subtract the nf from the measured spectral density $S[\Omega]$ and integrate the curve in the bandwidth of 500 Hz, to calculate the area under the power spectral density. The area calculated for each PZT drive was normalized by area in the case of undriven PZT, to calculate the effective temperature of the mechanical mode. From this, an effective temperature of $\sim$ 23,000 K was found at 1000 $\mu$V$^2/$Hz of white noise. The effective temperature varies linearly with the area under the power spectral density and is plotted in Fig. \ref{figure_3}. For comparison, a value of $g_0$ can be obtained by extracting the slope from the linear fit. The slope $\zeta$  is given by,
\begin{equation}{\label{equation2}}
	\zeta=\dfrac{m_\text{eff}\Omega_\text{m}^2\kappa^2}{2\pi\kappa_\text{ext}G^2n_\text{d}\hbar\omega_\text{c} k_\text{b}}
	\end{equation}
We put the value of slope $\zeta$ into equation \ref{equation2} giving $g_0/2\pi=$ 185$\pm5$ $\mu$Hz for this method again depending on 0.5 dBm of error made on calibration of the power input into the cavity, consistent with the previously derived value.

\subsection{\label{subsec:level3}OMIT/OMIA}
OMIT/OMIA is the optomechanical analogue of electromagnetically induced transparency observed in cold atoms \cite{fleischhauer2005electromagnetically}. It was theoretically predicted \cite{schliesser2009cavity} in 2009 and first observed in optical systems \cite{weis2010optomechanically,safavi2011electromagnetically}. Exploiting this effect could be useful to signal processing applications such as for delay lines or memories. Here we present the first measurement of OMIT/OMIA in a microwave cavity optomechanical system operated at room temperature. This measurement also enables us to extract $g_0$ with an additional independent characterization technique.

The cavity is driven by a strong pump at $\omega_\text{d}=\omega_\text{c}+\Delta$ and is probed with a relatively weaker signal at $\omega_\text{p}$ resulting in the radiation pressure force on the mechanical resonator at the difference frequency $\Omega=\omega_\text{p}-\omega_\text{d}$. The mechanics is coherently driven when $||\Omega|-\Omega_\text{m}| \lesssim \Gamma_\text{m}$. The driven mechanics generates sidebands on the pump tone. The sideband then interferes constructively or destructively with the probe tone, depending on the sign of $\Delta$, where $\Delta=\pm \Omega_m$, thus influencing the cavity transmission. When $\Delta=-\Omega_\text{m}$ (red pump), the sideband generated at $\omega_\text{c}$ interferes destructively with $\omega_\text{p}$ leading to OMIA. On the other hand, a transparent window is opened at $\omega_\text{c}$ (OMIT) when probing cavity transmission when $\Delta=\Omega_\text{m}$.    
\begin{figure}[ht!]
	\includegraphics[width=\columnwidth]{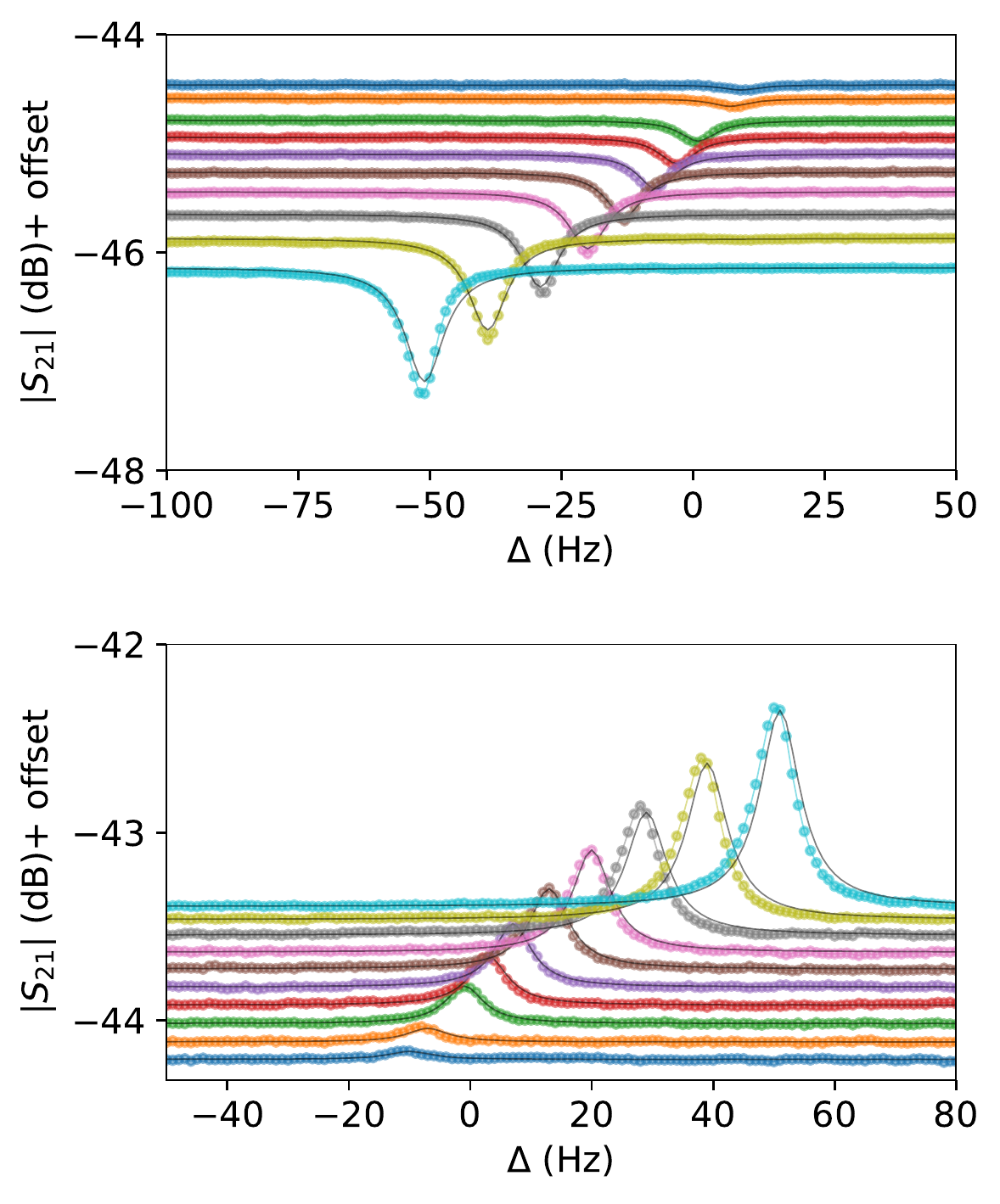}
	\caption{\label{figure_4} (top) OMIA was observed in the vicinity of $\omega_\text{c}$ when pump tone is applied at $\omega_\text{c}-\Omega_\text{m}$. The number of photons $n_\text{d}$ was varied from 1.7 $\times 10^{12}$ (blue curve) to 3.8 $\times 10^{13} $ (cyan curve). Grey lines are the fits. (bottom) OMIT was observed when pump tone was applied at $\omega_\text{c}+\Omega_\text{m}$. All the parameters are the same as the top plot.There is an offset in the y-axis for both the plots for clarity.
	}
\end{figure}

The circuit diagram for the measurement is discussed in appendix \ref{circuit diagram}. Fig. \ref{figure_4} (top) shows the OMIA response while using red pump tone at $\omega_\text{c}-\Omega_\text{m}$. The drive photons $n_\text{d}$ were varied from $1.7\times10^{12}$ to $3.8\times10^{13}$ photons. The weaker probe tone $\omega_\text{p}$ with -40 dBm of power applied by the source was swept in a narrow window of 200 Hz about $\omega_\text{c}$. Increasing the number of drive photons led to stronger driving of the mechanical resonator, thus increasing the amplitude of the absorption curve. Ideally, the absorption curves should be at $\omega_\text{c}$, but we observed a frequency shift in the absorption curves as we increased the number of photons shown in Fig. \ref{figure_4}. The y-axis in Fig. \ref{figure_4} is also offset for clarity. We observe a spring softening caused by the pump signal, which corresponds to a temperature elevation of the membrane of the order of 2 K at the maximum power. More information on this effect can be found in the appendix \ref{Spring softening}. The transmission $S_{21}[\omega_\text{p}]$ is given by \cite{kumar2022microwave},
\begin{equation}{\label{equation3}}
	S_{21}[\omega_\text{p}]=1-\dfrac{\chi_\text{c}[\omega_\text{p}]\kappa_\text{ext}/2}{1\mp g_0^2 n_\text{d} \chi_\text{c}[\omega_\text{p}]\chi_\text{m}[\omega_\text{p}]} \text{ ,}
\end{equation}
where $\chi_\text{c}^{-1}[\omega_\text{p}]=\dfrac{\kappa}{2}-i(\Omega+\Delta)$ and $\chi_\text{m}^{-1}[\omega_\text{p}]=\dfrac{\Gamma_\text{m}}{2}-i(\Omega\pm\Omega_\text{m})$. The upper (lower) sign in $S_{21}[\omega_\text{p}]$ and $\chi^{-1}[\omega_\text{p}]$ is for blue (red) pumping. Eq. \ref{equation3} was used to fit the curves in Fig. \ref{figure_4} extracting $g_0/(2\pi)$ = 192 $\pm$ 4  $\mu$Hz, consistent within $\sim$ 5\% of $g_0$ derived from direct power spectral density measurements detailed in subsection \ref{subsec:level1}. Similar measurements were made with blue pump at $\omega_\text{c}+\Omega_\text{p}$, finding a similar value of $g_0$ and we observing OMIT shown in Fig. \ref{figure_4} (bottom). 
\section{ Homodyne detection}
\begin{figure*}[ht!]
	\includegraphics[width=\linewidth]{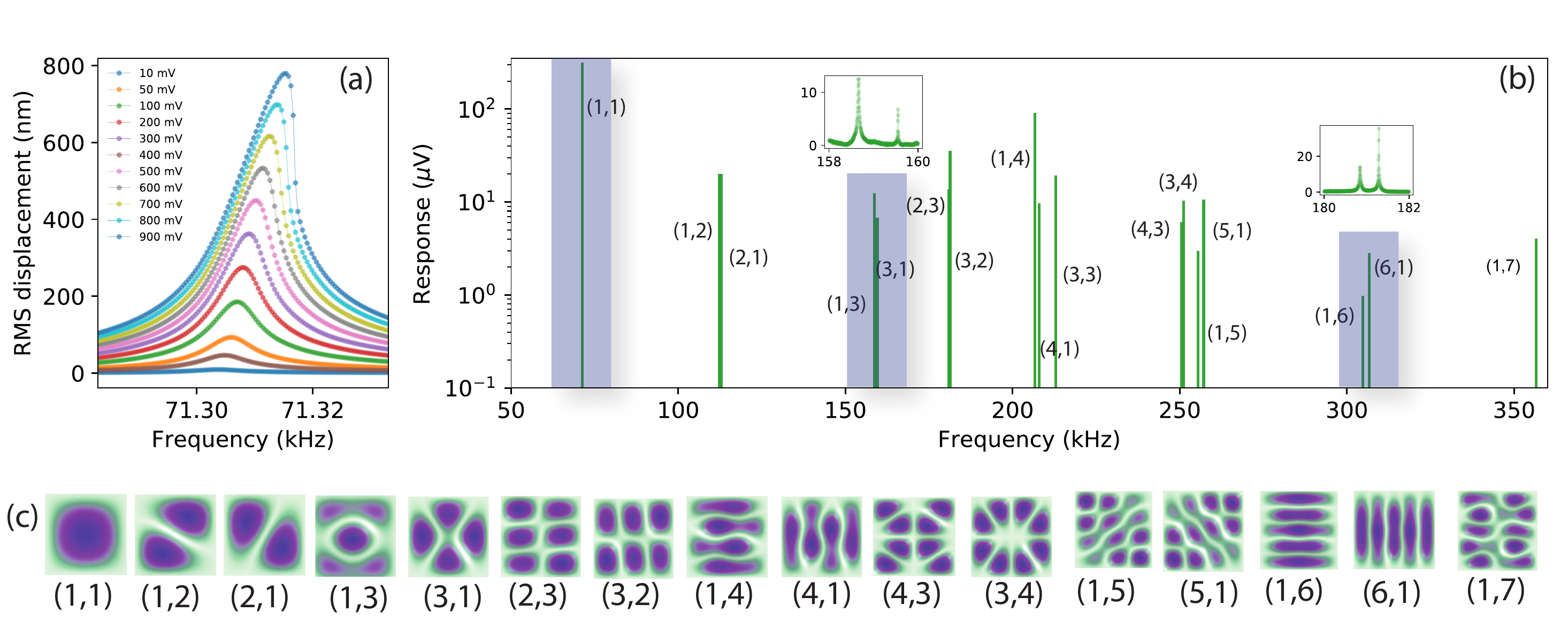}
	\caption{\label{figure_5} (a)  Non-linearity observed in mode (1,1) of the membrane by increasing the drive voltage from 10 mV$_\text{RMS}$ to 900 mV$_\text{RMS}$.
		b) Different eigenmodes of Si$_3$N$_4$ membrane resonator detected by sweeping the drive frequency $\Omega$ from 50 kHz to 400 kHz. c) Mode shapes for different eigenmodes of the membrane. Darker color shows the region of maximum amplitude 
	}
\end{figure*}

An alternative to direct detection with a spectrum analyzer is the homodyne detection method. A microwave interferometer type method, downconverting the signal at $\omega_\text{d}+\Omega_\text{m}$ by mixing with a local oscillator signal at $\omega_\text{d}$ to generate a signal at $\Omega_\text{m}$, measured by lock-in amplifier (see Fig. \ref{figure_8}). Here, $\omega_\text{d}=\omega_\text{c}$ is at the microwave resonance frequency, for a phase-sensitive measurement with maximum sensitivity. We measure the voltage spectral density $S_\text{V}[\Omega]$ of the signal at $\Omega_\text{m}$ using zoom fast Fourier transformation, where the signal at $\Omega_\text{m}$ is downmixed with a signal near $\Omega_\text{m}$, greatly increasing measurement speed. The voltage spectral density is given by,
\begin{equation}
	S_\text{V}[\Omega](\text{V}^2/\text{Hz})=  \dfrac{K\kappa_\text{ext}G^2 n_\text{d}}{\kappa^2}\hbar\omega_\text{c} S_\text{x} [\Omega] + \text{nf,}
\end{equation}
where $K$ is a scaling factor which depends on the input impedance of the lock-in amplifier, attenuation in the circuit and the phase difference of the microwave pump signal $\omega_\text{d}$ in different parts of the circuit. 
The square root of the area under the position spectral density $S_\text{x}[\Omega]$ shown in Fig. \ref{figure_2} gives the normalized RMS mechanical amplitude $x_\text{RMS}=\sqrt{\langle x \rangle^2} = 4 \text{ pm}$. The square root of the area under $S_\text{V}-\text{nf}$ at $\Omega_\text{m}$ is the RMS voltage $V_\text{RMS}$ measured by the lock-in amplifier such that
\begin{equation}\label{equ10}
V_\text{RMS}	=  \sqrt{\dfrac{K\kappa_\text{ext}G^2 n_\text{d}}{\kappa^2}\hbar\omega_\text{c}} x_\text{RMS.}
\end{equation}
 We subtracted the noise floor nf from $S_\text{V}$ and calculated the square root of the area under the curve giving $V_\text{RMS}=0.1$ $\mu\text{V}$. Putting the value $x_\text{RMS} = 4 \text{ pm}$ in Eq. \ref{equ10}, we extracted 
 \begin{equation}
 	K_1	=  \sqrt{\dfrac{K\kappa_\text{ext}G^2 n_\text{d}}{\kappa^2}\hbar\omega_\text{c}} = 4 \times 10^4 \text{,}
 \end{equation}
 where $K_1$ is the scaling factor to convert $V_\text{RMS}$ measured by the lock-in to the RMS displacement $x_\text{RMS}$  of the mechanical resonator.
  By applying an AC electrical signal on the PZT at $\Omega$ it will then impart a sinusoidal force at $\Omega$ to the mechanical resonator, driving it coherently. Measurement circuit is shown in appendix \ref{circuit diagram}.  
The eigenfrequencies of the mechanical resonator were characterized by sweeping $\Omega$ from 50 kHz to 400 kHz, and measuring response at the lock-in at the swept frequency. The different eigenmodes are shown in Fig. \ref{figure_5}. The eigenfrequencies of different modes match with the theoretical values within 1 \%. Ideally, degenerate pairs of modes like (2,3),(3,2); (1,6),(6,1), etc. should have identical eigenfrequencies, but we observed a mismatch shown in the inset of Fig. \ref{figure_5}. This mismatch can be attributed to the unsymmetrical edges of the square membrane.

Characterizing nonlinearity in the mechanical resonator is also an important step with its applications in signal processing and precision sensing \cite{zhou2021high}. For this characterization, we measured the response of the fundamental mode (1,1) of the Si$_3$N$_4$ membrane. The PZT was driven by an upward sweeping sinusoidal signal across the mechanical resonance. The response of the mechanical resonator can be modeled using a duffing non-linear equation of motion,
\begin{equation}
\dfrac{\partial^2 x}{\partial t^2}+\Gamma_\text{m}\dfrac{\partial x}{\partial t}+\Omega_\text{m}^2 x + \alpha x^3=\dfrac{F\cos(\Omega t)}{m_\text{eff,11}}
\end{equation}
where $x$ is the mechanical displacement, $\alpha$ is the duffing parameter. The drive voltage applied to the PZT varied from 10 mV$_\text{RMS}$ to 900 mV$_\text{RMS}$. The response of the mechanical resonator is shown in Fig. \ref{figure_5}. For weaker drives, the response of the mechanical resonator is Lorentzian around the resonance frequency $\Omega_\text{m}$. Increasing the drive voltage, the maximum amplitude $x_\text{max}$ shifts to higher frequencies $\Omega_\text{eff}$ visible in the Fig. \ref{figure_5}. This behavior is well approximated by the backbone curve equation as \cite{zhou2021high},
\begin{equation}
	x^2_\text{max}=\dfrac{8}{3}\dfrac{\Omega_\text{m}}{\alpha}(\Omega_\text{eff}-\Omega_\text{m})
\end{equation} 
Using the above equation we extracted the duffing parameter $\alpha=1.68\times 10^{19} $ m$^{-2}$ s$^{-2}$. We also theoretically determined the duffing parameter using the expression given by Cattiaux \textit{et al.} \cite{cattiaux2020geometrical} and found it to be 4.2 $\times 10^{19} $ m$^{-2}$ s$^{-2}$ which closely matches with experimental value. In conclusion, with this architecture, we can probe a displacement amplitude from 4 pm to 100 nm in the linear regime, up to hundreds of nm, and more in the non-linear regime.
\subsection{Calibrated Homodyne}
We have used the phase sensitive homodyne detection technique to measure the optomechanical sidebands of the membrane-cavity system. However, to calculate a $g_0$ using this technique a calibration of the circuit is required. We wish to avoid a direct calibration of the greater microwave circuit's transmission, which will slowly drift. Instead we create a reference sideband by phase modulating the source's microwave signal, at $\omega_\text{d}$, by a known magnitude $\phi_\text{mod}$ and frequency $\Omega_\text{mod}$, close to $\Omega_m$. This method follows the methodology of Ref. \cite{gorodetksy2010}, but is adapted from optics to microwaves using phase noise measurement techniques \cite{packard1985}. Following Ref. \cite{gorodetksy2010} we derive an equation for the vacuum optomechanical coupling in the strongly sideband unresolved regime,
\begin{equation}
	g_0^2 = \left[ \kappa_\text{L} \phi_\text{mod} \sin(\psi_\text{r}/2) \right]^2 \frac{ \Gamma_\text{m} S_\text{V}[\Omega_\text{m}]}{16 \text{BW} \langle n_\text{m} \rangle S_\text{V}[\Omega_\text{mod}]} \text{ ,}
\end{equation}
where $\psi_\text{r}$ is the relative phase difference between the two arms of the circuit, BW is the bandwidth of the lock-in measurement, and $\langle n_\text{m} \rangle$ is the average occupation of the mechanical mode. Then $S_\text{V}[\Omega_\text{m}]$ and $S_\text{V}[\Omega_\text{mod}]$ are the peak voltage noise values of the optomechanical and modulated signal respectively, as measured by the lock-in. $\psi_\text{r}$ is approximately the product of the relative microwave time delay $\tau_\text{r}$ between the two arms and $\Omega_m$, for our measurement setup this is calculated to be $\psi_\text{r} = 0.686$ mrad. In an earlier experiment we used this technique to measure a coupling for the same membrane-cavity system of $g_0/(2\pi) = 115 \; \mu$Hz, with the reduced value likely from an unoptimised microwave cavity coupling leading to a reduction in sensitivity. Though this method requires additional setup, it does eliminate the need for sensitive microwave detection equipment.
\section{Conclusions}
To summarize, we have demonstrated a cavity optomechanical system realized using a pre-stressed Si$_3$N$_4$ membrane coupled to a re-entrant microwave cavity made of Cu. Our microwave optomechanical setup is capable of resolving the thermomechanical motion of the Si$_3$N$_4$ membrane at RT. Moreover, the PZT attached to our setup helped us to excite different mechanical modes of the membrane and also deduce the geometric non-linearity. We extracted the single photon coupling rate $g_0/(2\pi)=$ 188.5 $\pm$ 5 $\mu$Hz for the fundamental mode using direct spectral and two-tone measurements and cooperativity of $C=4g_0^2n_\text{d}/(\kappa\Gamma_\text{m})$ = 1.2 $\times 10^{-3}$ at the maximum pump power used which is close to 13 dBm going into the cavity. The dynamic optomechanical effects such as cooling and heating of the mechanical mode are within our reach just by improving upon the architecture of our setup. By precision machining of our cavity, we can further reduce the gap between the membrane and the top of the pillar of the cavity to less than 10 $\mu$m, leading to 10 times increase in $g_0$, therefore reaching the self oscillation regime with $n_\text{d} \sim 10^{14}$ in the cavity at RT. Further, we can develop microwave notch filters and amplifiers using the OMIT/OMIA. Using a strong pump tone at $\omega_\text{d}=\omega_\text{c}+\Delta$ with $\Delta = \Omega_\text{m}$ and using a weaker probe tone at $\omega_\text{p}$ to sweep the response of the cavity, we can achieve a gain at $\omega_\text{c}$ close to 25 dB with $n_\text{d} \sim 10^{12}$ in the cavity with the gap $x \sim 10$ $\mu$m. This magnitude of gain at RT is comparable to similar measurements made at cryogenic temperatures \cite{massel2011microwave}. On the other hand, notch filters can be realized using the two tone measurement but with $\Delta = - \Omega_\text{m}$. The novelty of our microwave optomechanical system to work at RT and the prospects of making amplifiers, filters and notch filters with performances similar to the systems working at cryogenic temperatures is exciting and is within our reach. Recently, Serra \textit{et al.} \cite{serra2021silicon} mentioned the importance of isolating the mechanical resonator from the substrate to improve on the quality factor and a step towards room temperature quantum optomechanics. Our system is flexible to quickly test different kind of mechanical resonators (trampoline, phononic crytals, SiN membranes with soft clamping, etc.). Also, using the two pillars inside the cavity \cite{goryachev2015creating} and coupling them to two or more mechanical resonators, which our system is capable of, provides an ideal platform for multimode optomechanics. 

\begin{acknowledgments}
This research is supported by the Royal Society (UF150140, RGF$\backslash$EA$\backslash$180099, RGF$\backslash$R1$\backslash$180059, RGF$\backslash$EA$\backslash$201047, RPG$\backslash$2016$\backslash$186, URF$\backslash$R$\backslash$211009), the EPSRC (EP/R04533X/1), and the STFC (ST/T005998/1). For the purpose of open access, the author has applied a Creative Commons Attribution (CC BY) licence (where permitted by UKRI, ‘Open Government Licence’ or ‘Creative Commons Attribution No-derivatives (CC BY-ND) licence’ may be stated instead) to any Author Accepted Manuscript version arising.
\end{acknowledgments}

\appendix
\section{\label{gap_estimation}Estimating gap between re-entrant post and the Si$_3$N$_4$ membrane}
\begin{figure}[ht!]
	\includegraphics[width=\columnwidth]{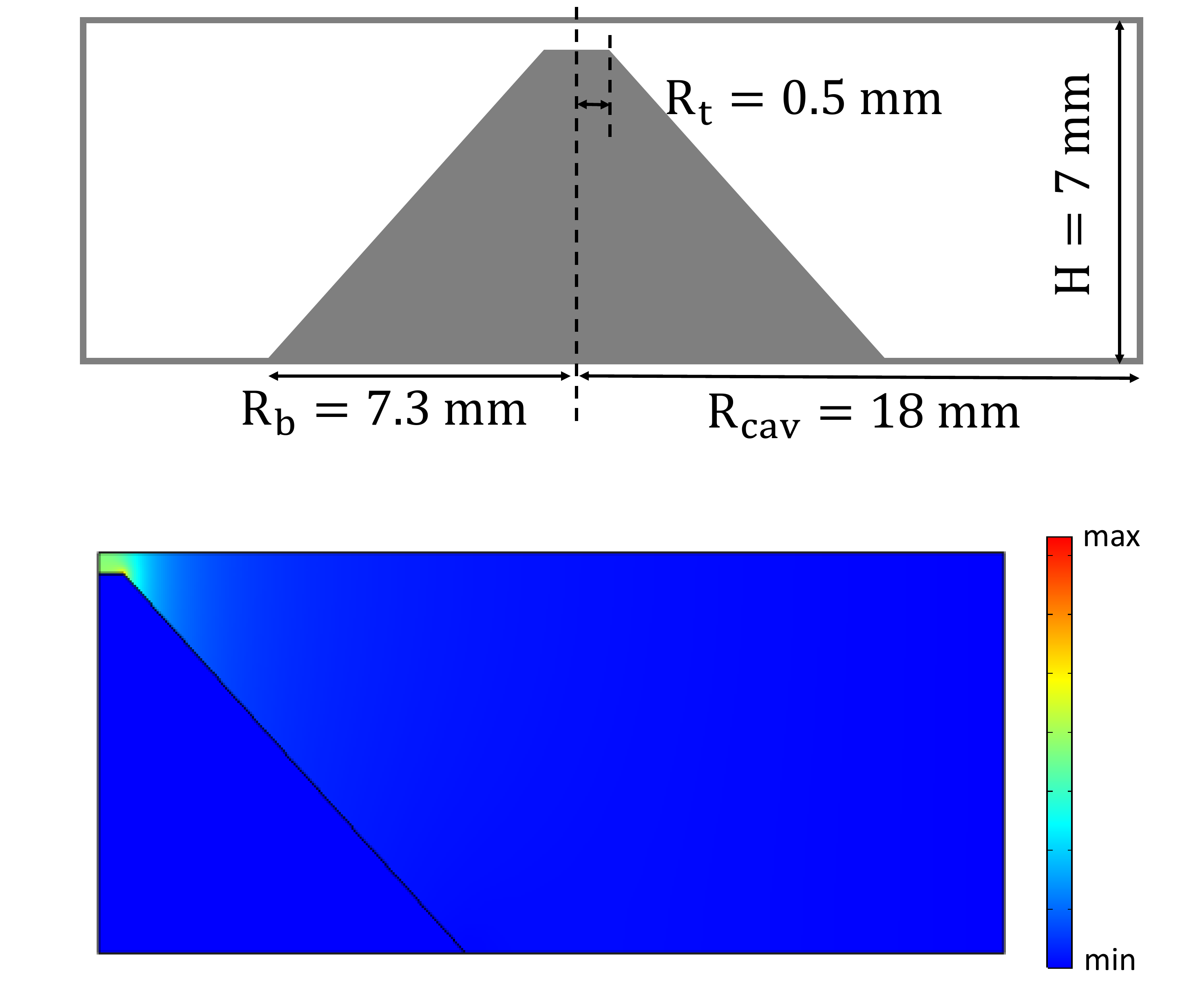}
	\caption{\label{figure_6} (top) Figure showing the dimensions of the simulated re-entrant cavity, (bottom) FEM simulation of the re-entrant cavity where the colour scale indicates the magnitude of the electric field in the cavity. The resonance frequency predicted by simulation is 4.62 GHz. }
\end{figure}

The microwave cavity is designed with a cone-shaped pillar inside it. The cavity of radii $R_\text{cav}=$ 18 mm and height $H=$ 7 mm consists of a cone-shaped pillar with top radius of 0.5 mm and the bottom radii of 7.3 mm shown in Fig. \ref{figure_6}. The resonance frequency of the microwave cavity primarily depends on the gap between the top part of the cone and the chip. We performed FEM simulation using COMSOL shown in Fig. \ref{figure_6} to extract the microwave resonance frequency. With a gap of 100 $\mu$m between the lid/chip and the top radii of the cone, the simulated resonance frequency of 4.62 GHz matches with the experimental value of the resonance frequency within 1 \%.
\section {\label{Spring softening}Spring softening of Si$_3$N$_4$ membranes}
In subsection \ref{subsec:level3}, we discussed the spring softening of the Si$_3$N$_4$ membrane. We observed a decrease in the resonance frequency of the mechanical mode associated with an increase in the applied pump power. The magnitude of the frequency shift is of the order of 50 Hz at the maximum input power used in this experiment (13.5 dBm). This is observed at zero detuning ($\omega_\text{d}=\omega_\text{c}$), and for both blue-detuned ($\omega_\text{d}=\omega_\text{c}+\Omega_\text{m}$) and red-detuned  ($\omega_\text{d}=\omega_\text{c}-\Omega_\text{m}$) pumps. In the strongly unresolved sideband regime ($\kappa>>\omega_\text{m}$), the drive photon number is nearly the same in all three cases (green, blue and red pumping).
At the highest input power $P_\text{in}= $13.5 dBm (0.022 W), the drive photon number in the cavity is $n_\text{d}=3.8\times10^{13}$. Hence, the total power dissipated by the cavity is given by
\begin{equation}
    P_{\rm diss}=(\hbar\omega_{\rm d}) n_{\rm d} \kappa_{\rm int} \simeq 4\ \mathrm{mW}
\end{equation}
with $\kappa_{\rm int} = \kappa - \kappa_{\rm ext} = (2\pi) \times 5.502$ MHz. We note that the internal cavity loss has a contribution from the electrical resistance of the thin gold layer coating of the chip. Hence, a fraction of the electrical power is dissipated in the vicinity of the chip, which can cause a local heating of the membrane.

\begin{figure}[!ht]
	\includegraphics[width=8 cm]{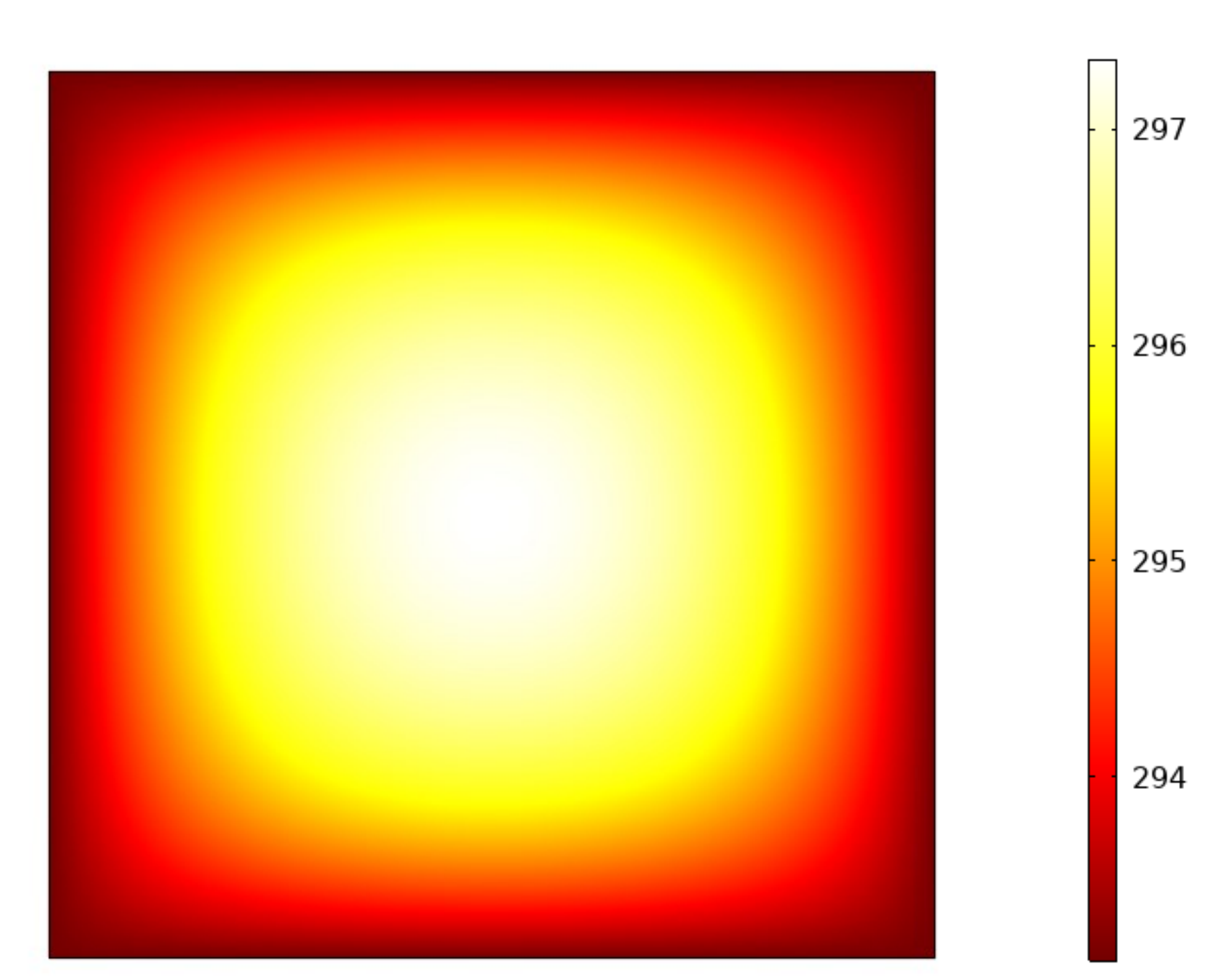}
	\caption{\label{figure_9} (top) 2D FEM simulation using COMSOL showing the temperature map of 2 mm $\times$ 2 mm Si$_3$N$_4$ membrane. The boundaries are acting as heat sink at RT.  }
\end{figure}

Using COMSOL, we showed in a simple 2D FEM simulation, that a local heating effect of the membrane corresponding to a fraction of the total dissipated power, of the order of a percent, can explain on an average temperature elevation of the membrane $\Delta T= 2$ K. In this model, we considered a 2 $\times$ 2 mm free-standing Si$_3$N$_4$ membrane, and assumed a perfect thermalization of the membrane's boundary at the ambient room's temperature. Fig.\ref{figure_9} shows the membrane's temperature profile. 

Because of thermal expansion, the membrane's temperature elevation causes a resonant mode frequency shift. Using the expression given by Zhang \textit{et al.} \cite{zhang2020radiative}
\begin{equation}
	\Delta f \approx -\dfrac{\alpha E f_\text{11}}{2\sigma(1-\nu)}\Delta T_\text{m} \text{ ,}
\end{equation}
one can calculate the frequency shift $\Delta f$, where $\alpha=2.2 \times 10^{-6} \text{ K}^{-1}$ is the thermal expansion coefficient, $E=250$ GPa \cite{Norcada} is the Young's modulus, $\sigma = 1$ GPa \cite{Norcada} is the in-built stress and $\nu=$ 0.23 is the Poisson's ratio of the silicon nitride membrane. The frequency shift calculated using this model is consistent with a power absorbed by the membrane of the order of $1 \%$ of the total dissipated power, which seems realistic. An in depth description of this thermal heating effect would fall outside the scope of this paper.

\section {\label{circuit diagram}Circuit diagrams}
This appendix will discuss the circuit diagrams used for each distinct measurements. Fig. \ref{figure_7} shows the circuit diagram for thermal noise measurements, white noise drive measurements, and OMIT/OMIA. 
\subsection{Thermal noise measurements}
\begin{figure*}[ht]
    \includegraphics[width=17cm]{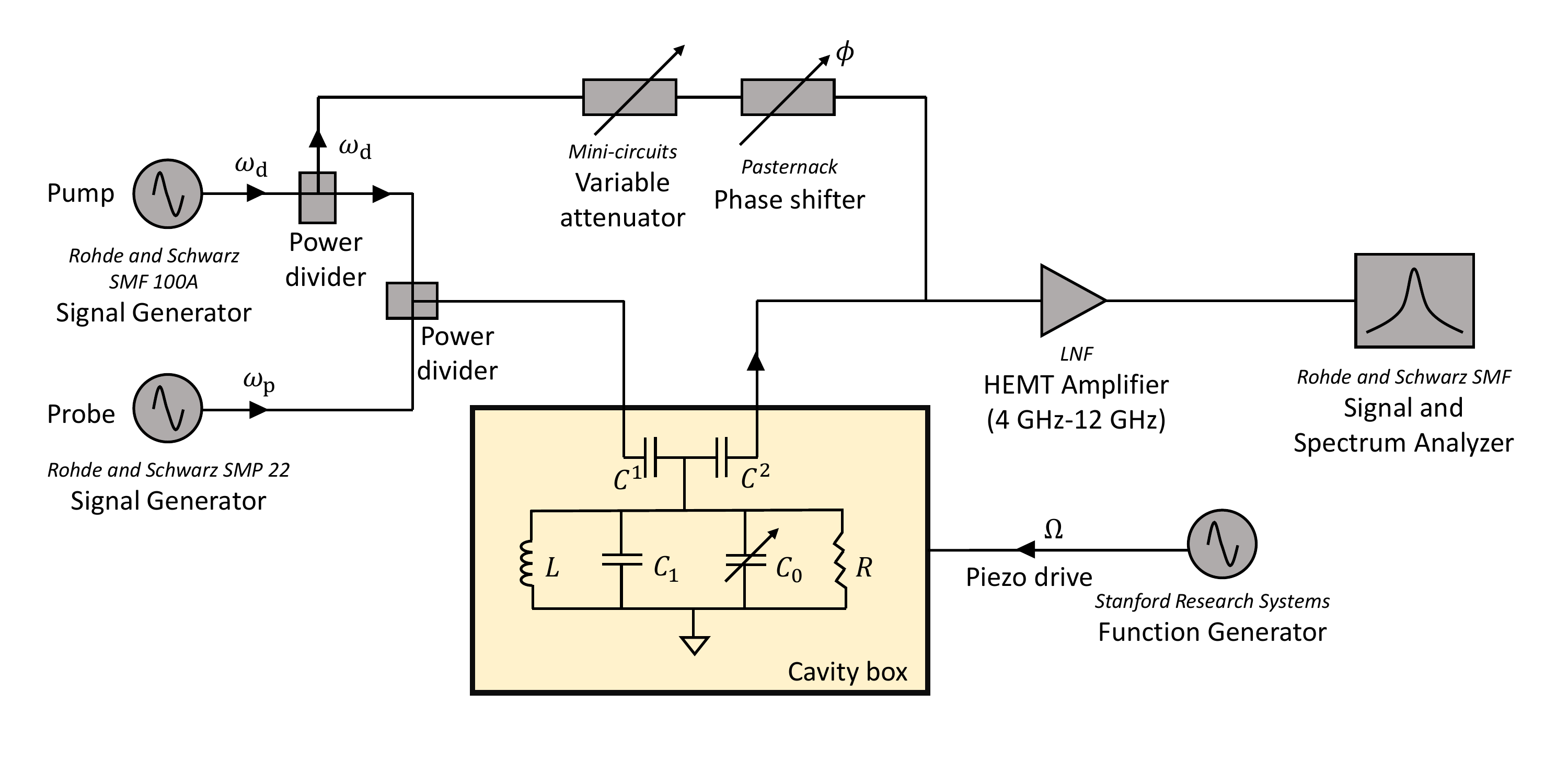}
	\caption{\label{figure_7} Measurement circuit diagram for direct thermal noise spectral measurements, white noise driven spectral measurements and two-tone measurements. The signal was measured using the spectrum analyzer.}
\end{figure*}

For the thermal noise measurements, the piezo drive and probe tone were off. The microwave pump signal at $\omega_\text{d}$ generated by the microwave source (Rohde and Schwarz SMF 100A signal generator) was fed to the re-entrant cavity via port 1 shown in shaded light yellow in Fig. \ref{figure_7}. The re-entrant cavity can be modeled as a parallel RLC circuit. The capacitance $C_0$ between the end post and the mechanically compliant Si$_3$N$_4$ membrane resonator is then varied by the oscillations of the membrane. Due to thermomechanical motion of the mechanical resonator with resonance frequency $\Omega_\text{m}$, the signal at $\omega_\text{d}$ is modulated such that the signal coming out of the cavity from port 2 has three frequency components ($\omega_\text{d}$, $\omega_\text{d}+\Omega_\text{m}$, $\omega_\text{d}-\Omega_\text{m}$). We cancel the signal coming out of port 2 at $\omega_\text{d}$ by combining it with another signal at $\omega_\text{d}$ by adjusting the phase shifter (Pasternack Phase shifter) ($\phi$) and the variable attenuator (Mini-Circuits digitally controlled variable attenuator). The signal is then amplified using a room temperature amplifier (LNF RT HEMT, 4 GHz-12 GHz) before measuring the noise using the spectrum analyzer (Rohde and Schwarz SMF Signal and Spectrum Analyzer). The pump tone was applied at the microwave resonance frequency such that $\omega_\text{d}=\omega_\text{c}$, where $\omega_\text{c}$ is the microwave resonance frequency. We used 5 dBm of input power from the generator, which was further attenuated by power dividers and coaxial lines such that the power reaching the input port of the cavity was -6.8 dBm which was measured using a Keysight microwave power meter. The gain in the circuit was 23.5 dBm. We then measured the signal at $\omega_\text{c}+\Omega_\text{m}$ using the spectrum analyzer. 

\subsection{White noise driven spectral measurements}
For the white noise driven Si$_3$N$_4$ membrane, the piezoelectric transducer was excited by white noise in the bandwidth of 10 MHz using function generator (Stanford SRS DS345), keeping every element of the circuit the same as in the thermal noise measurments. The signal was measured using the spectrum analyzer.

\subsection{Two-tone measurements}
When measuring the transmission of the cavity using two-tone, the piezo drive was turned off. The pump tone at $\omega_\text{d}$ was combined with a weaker probe tone at $\omega_\text{p}$ generated by another microwave source (Rohde and Schwarz SMP 22 Signal generator) before the signal was fed to the microwave cavity. The pump tone at $\omega_\text{d}$ was kept fixed such at $\Delta=\omega_\text{d}-\omega_\text{c}$. The weaker probe tone with power of -40 dBm at $\omega_\text{p}$ was swept from $\omega_\text{c}-2\pi\times100$ to  $\omega_\text{c}+2\pi\times100$. The pump tone power generated by the source was also varied from 16 dBm to 25 dBm. 

\subsection{Homodyne detection}
\begin{figure*}[hb]
	\includegraphics[width=17cm]{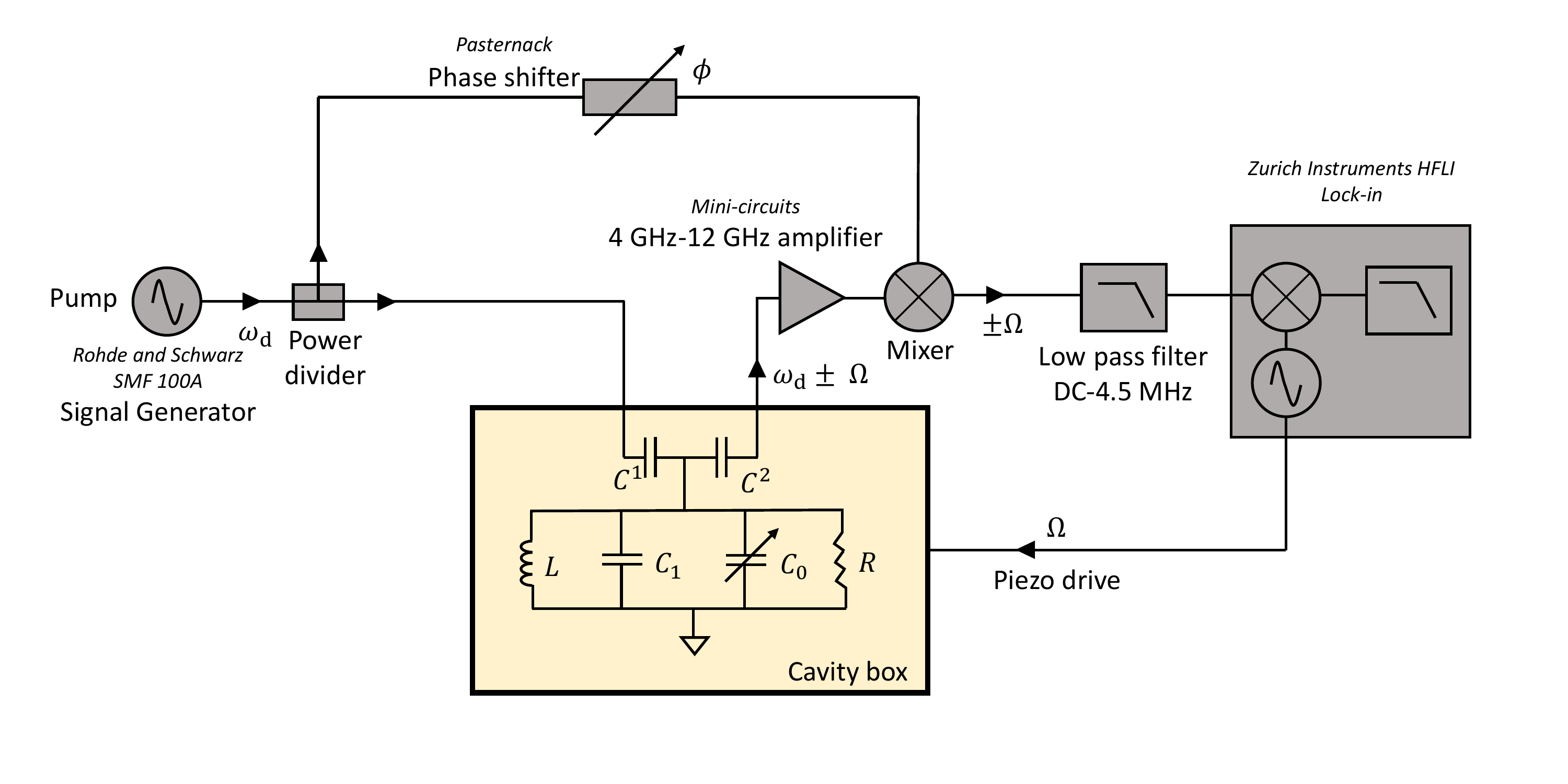}
	\caption{\label{figure_8} Measurement circuit diagram for Homodyne detection scheme. We used this circuit to measure the response of the periodically driven mechanical resonator and characterize the geometric non-linearity.}
\end{figure*}

For the homodyne detection scheme, we use lock-in amplifier to detect the signal. The microwave pump at $\omega_\text{d}$ was again fed to the re-entrant cavity via port 1 as shown in Fig. \ref{figure_8}. In case of driven periodic motion of the membrane by exciting the piezo at an RF frequency $\Omega$ generates sidebands on the pump signal at $\omega_\text{d}\pm \Omega$. The transmitted signal is amplified using an RT microwave amplifier (Mini-circuits microwave amplifier, 4 GHz-12 GHz). The amplified signal is then mixed with a microwave signal at $\omega_\text{d}$ such that the signal coming out of the mixer is at $\pm \Omega$. We cancel the DC component by adjusting the phase shifter also shown in the Fig. \ref{figure_8}. The RF signal is filtered removing any high frequency component before being measured by the lock-in amplifier (Zurich Instrument, HF2LI). The RF signal $\Omega$ driving the piezo is swept across the resonance frequency of the membrane to measure the response of the same and also characterize the geometric non-linearity. In case when the piezo is turned off, the thermomechanical motion will generate sidebands on the pump signal at $\Omega_\text{d}\pm \Omega_m$. The transmitted signal is then amplified and downconverted before measuring the signal with a lock-in amplifier using the zoom FFT method.

\clearpage
\bibliography{aipsamp}

\end{document}